\documentclass[authoryear,preprint,12pt]{elsarticle}

\usepackage{algorithm}
\usepackage{algorithmic}
\usepackage{caption}
\usepackage{graphicx} 
\usepackage{subcaption}
\usepackage{multirow}
\usepackage{tabularx}
\usepackage{mathtools}
\usepackage{colortbl}
\usepackage{hyperref}
\usepackage{footnote}
\usepackage{xcolor}
\usepackage[T1]{fontenc}

\usepackage{amssymb}

\newcommand{\citeay}[1]{%
  \citeauthor{#1} \citep{#1}%
}

\hypersetup{
  colorlinks=true,     
  linkcolor=blue,      
  citecolor=green,     
  filecolor=magenta,   
  urlcolor=cyan        
}

\journal{ArXiv}

\begin{document}

\begin{frontmatter}

\title{Adaptive Asynchronous Work-Stealing for distributed load-balancing in heterogeneous systems}

\author[1]{João B. Fernandes\corref{cor1}}
\ead{joao.batista.fernandes.094@ufrn.edu.br}
\affiliation[1]{organization={Dept. de Engenharia de Computação e Automação, UFRN}, country={Brazil}}
\cortext[cor1]{Corresponding author}

\author[2]{Ítalo A. S. de Assis}
\ead{italo.assis@ufersa.edu.br}
\affiliation[2]{organization={Dept. de Engenharia e Tecnologia, UFERSA},
           country={Brazil}}

\author[3]{Idalmis M. S. Martins}
\ead{idalmismilian@bct.ect.ufrn.br}
\affiliation[3]{organization={Escola de Ciência e Tecnologia, UFRN},
           country={Brazil}}

\author[1]{Tiago Barros}
\ead{tbarros@dca.ufrn.br}

\author[1]{Samuel Xavier-de-Souza}
\ead{samuel@dca.ufrn.br}

\begin{abstract}
Supercomputers have revolutionized how industries and scientific fields process large amounts of data. These machines group hundreds or thousands of computing nodes working together to execute time-consuming programs that require a large amount of computational resources. Over the years, supercomputers have expanded to include new and different technologies characterizing them as heterogeneous.
However, executing a program in a heterogeneous environment requires attention to a specific aspect of performance degradation: load imbalance. 
In this research, we address the challenges associated with load imbalance when scheduling many homogeneous tasks in a heterogeneous environment. To address this issue, we introduce the concept of adaptive asynchronous work-stealing. This approach collects information about the nodes and utilizes it to improve work-stealing aspects, such as victim selection and task offloading. Additionally, the proposed approach eliminates the need for extra threads to communicate information, thereby reducing overhead when implementing a fully asynchronous approach. Our experimental results demonstrate a performance improvement of approximately 10.1\% compared to other conventional and state-of-the-art implementations.
\end{abstract}


\end{frontmatter}

\section{Introduction}
\setcounter{page}{1}
\label{intro}

    Supercomputers are high-performance machines with hundreds or thousands of computing nodes working together. They have expanded and incorporated new and diverse technologies. This includes the addition of different core types, different memory configurations, and the use of co-processors such as Graphics Processing Units (GPUs) in some compute nodes. As a result, supercomputers have become heterogeneous, particularly regarding each node's processing speed and architecture-related features. In this scenario, numerous fields of science and industry, such as seismic analysis and meteorology, rely heavily on supercomputers to process vast amounts of data. Therefore, efficiently using all available computational resources, whether heterogeneous or not, is highly desirable to maximize utilization to the largest possible extent.     
    	
    Executing computational jobs in a heterogeneous environment has become increasingly desirable. Consequently, job schedulers such as SLURM ~\citep{slurm2003}, widely employed in supercomputers, have incorporated heterogeneous job support~\citep{Slurm2022}. This enhancement aligns with the increasing requirement for executing diverse job types within the system. Cloud computing is another example where running programs with different node types can be beneficial, especially using pre-emptive instances, which are cheaper to use but have limited availability per type. 

    To achieve optimal execution of parallel programs in a heterogeneous environment, load balance is the performance aspect of the most importance. Otherwise, load imbalance can occur when the workload of a program is divided into processes and distributed among different processing units with varying performance. Faster processes may wait for slower ones to complete their execution, leading to inefficient resource utilization. Several factors can cause load imbalance, including data distributions, conditional branching within the program, performance differences among the nodes involved in the execution, or how tasks are distributed among the processes. 

    Our aim in this work is to enhance the performance of distributed applications on heterogeneous supercomputers. To achieve this, we propose a novel scheduling approach designed to minimize communication overhead and enable efficient program scaling across many heterogeneous node types, namely adaptive asynchronous work-stealing (A2WS). 

    As a case study, we present performance improvements on an acoustic seismic modeling algorithm running in heterogeneous environments. This application is widely used in the oil and gas industry to analyze and image the Earth's subsurface. To improve performance, the evaluation of a large set of shots might be distributed across multiple computational nodes of a large-scale supercomputer.

    These are the main contributions of this work:

    \begin{itemize}
        \item The proposed work-stealing algorithm introduces an asynchronous global information propagation method based on local interaction and one-sided MPI communication 
        \item We propose an adaptive task-stealing method that uses the locally collected information to optimally adjust the number of tasks to be stolen among processing nodes
        \item The proposed victim selection process minimizes the number of steals by considering the collection information and matching the best pair of victims and the thief. 
        \item We propose a fully asynchronous theft mechanism that enables a more flexible and efficient task redistribution process.      
        \item We proposed a proactive approach that allows the work-stealing to happen anytime after the first task is finished, avoiding empty queues.
        \item We compared the A2WS method and other load-balancing techniques described in the literature for distributed systems.
    \end{itemize}

    The structure of this work is as follows. Section~\ref{sec:wave} provides a brief overview of seismic modeling and discusses the key implementation aspects relevant to our scheduler. Section~\ref{sec:a2ws} presents a detailed explanation of A2WS, highlighting its distinctive features. Section~\ref{sec:results} presents the results obtained from our comparative experiments between A2WS and the literature methods. In Section~\ref{sec:related}, we compare our scheduling method with other works found in the related literature, emphasizing its original aspects. Finally, Section~\ref{sec:conclusions} summarizes our conclusions regarding the proposed method.
      
\section{Adaptive Asynchronous Work-Stealing}
\label{sec:a2ws}

    In parallel programming, work-stealing is a scheduling method originally designed for shared-memory applications~\citep{Blumofe1999}. In work-stealing, each process maintains a queue of tasks; in recent implementations, a deque is used~\citep{Chase2005}. Unlike traditional queues, a deque structure allows for adding and removing tasks from both the front (head) and the back (tail). New tasks are initially added to the head of the deque. When a process needs a new task, it takes it from the head of the same deque. When a processor exhausts its tasks, it searches for another process and steals tasks from the tail of its deque. The process that initiates the stealing process is known as the \emph{thief}, while the target process of the stealing is called the \emph{victim}.
    
    In a distributed system, thieves must continually select victims and steal tasks from them. Selecting the victim randomly is the most common approach because it demonstrates good performance and is easy to implement~\citep{Arora1998}. However, steal failures can significantly slow down work-stealing performance~\citep{Perarnau2014}. Another critical factor is the number of transferred tasks, which is directly related to the frequency of steal attempts. 
    When the steal operations involve transferring only a few tasks or less than necessary, the number of steals required to maintain load balancing within the program tends to increase. Moreover, if more tasks are transferred than actually needed, the work-stealing can lead to an escalation in stealing, as another process may have to steal these tasks that have already been stolen. The amount of stealing substantially impacts performance because it adds communication overhead in distributed systems.

    Our work-stealing proposal gathers information about each process and uses it to adapt the work-stealing scheduling to the underlying environment. Algorithm~\ref{alg:a2ws} presents how the proposed approach performs. The information is updated in Lines~\ref{l:a2ws:up1} and \ref{l:a2ws:up2}. The exchange of information among the processes occurs through a function call, as shown in Line~\ref{l:a2ws:ic}. Every process has an information vector that includes the total number of tasks, the average task runtime for each process, and a respective flag to indicate if the information needs to be communicated. The function \textbf{info\_communication()} is responsible for checking if any of these flags are \textit{true}, and only in such cases, the process sends its local information to the neighboring nodes. After sending the needed information, all flags are reset to \textit{false}. Only new information is exchanged to avoid unnecessary communication.

    The information collected serves three purposes: to compute the number of tasks to be transferred (steal rate), select the victim, and anticipate the theft (preemptive stealing). The steal rate ($S$) is calculated using the \textbf{steal\_equation()} function in Line~\ref{l:a2ws:se} in Algorithm~\ref{alg:a2ws}. We use this value along with the information collected from the other processes to identify a suitable victim (Line~\ref{l:a2ws:sv}). If a victim is found, we initiate the stealing process (Line~\ref{l:a2ws:st}). Once the stealing is completed, the process retrieves a task from the deque (Line~\ref{l:a2wsgt}) and concludes the loop iteration. 

    \begin{algorithm}
    \caption{A2WS algorithm.}
    \label{alg:a2ws}
    \begin{algorithmic}[1]
        \WHILE{ the process has task }
            \STATE update\_process\_info() \label{l:a2ws:up1}
            \IF{ the process ran a task} \label{l:a2ws:if}
                \STATE $S \gets$ steal\_equation() \label{l:a2ws:se}
                \STATE $v \gets$ select\_victim($S$) \label{l:a2ws:sv}
                \IF{exist $v$}
                    \STATE steal\_task($v$, $S$) \label{l:a2ws:st}
                \ENDIF
            \ENDIF
            \STATE $T_{id} \gets$ get\_task\_id() \label{l:a2wsgt}
            \STATE update\_process\_info() \label{l:a2ws:up2}
            \STATE execute($T_{id}$)
            \STATE info\_communication() \label{l:a2ws:ic}
        \ENDWHILE
    \end{algorithmic}
    \end{algorithm}
        
    The proposed general approach shown in Algorithm~\ref{alg:a2ws} incorporates the following features that address the challenges of efficiently executing workloads in distributed heterogeneous environments, optimizing the load balancing and network usage: smart stealing, limited information communication, and asynchronous theft. These features work together to optimize load balancing and network utilization in a heterogeneous environment, improving the system's overall performance.
    
    Smart stealing may reduce the need for excessive stealing, thereby decreasing network usage. Simultaneously, asynchronous theft may increase network demands due to concurrent stealing activities, but its preemptive feature may distribute this demand over the execution. At the same time, we prevent excessive overhead by limiting the propagation of information. 

    The following subsections present the details of each of the three key features of our proposal. 
\subsection{Limited Information Communication}
\label{subsec:infoComm}

    To reduce the overhead of exchanging information among processes in A2WS, we propose communicating information as a bidirectional ring network. 
    Compared to traditional broadcast communication, the ring approach can reduce overhead while gradually updating all nodes with sufficiently accurate information. 
    To enhance efficiency and avoid synchronization issues, we leverage RDMA (Remote Direct Memory Access) \citep{MPIRDMA2018} operations provided by MPI \citep{MPI1994} methods such as \textit{Get} and \textit{Put}. RDMA enables direct memory access between processes without synchronization, further optimizing the communication process in A2WS.

    Periodic communication within the ring network is essential to ensure that all processes stay updated with the information vector. When multiple processes attempt to write to the same positions within this vector, it becomes necessary to establish mutual exclusion regions to prevent race conditions, which generate synchronization overhead. To mitigate this overhead, we employ a strategy of partitioning the vector into two distinct regions. In this approach, each process, denoted as $p_i$, transmits information with indices $j > i$ to process $p_{i-1}$ and information with indices $j < i$ to process $p_{i+1}$. This isolation of write operations effectively eliminates the need for a mutual exclusion region.
    
\begin{figure}
        \centering
        \includegraphics[width=.5\textwidth]{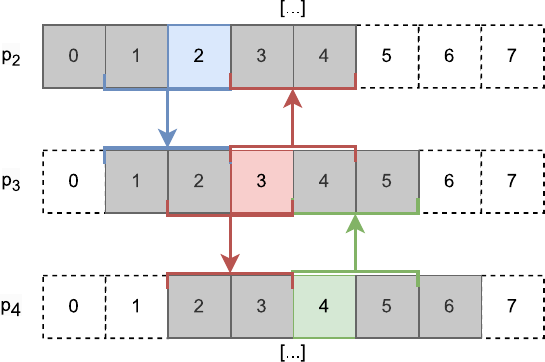}
        \caption{Diagram exemplifying the information-communication of $3$ processes ($p_2$, $p_3$, $p_4$) of a system with $8$ processors and $R = 2$. The arrows are representations of the MPI Put operation. Therewith, we can see all possible operations on the $p_3$ information vector.}
        \label{fig:info-exp}
    \end{figure}

    However, it is important to acknowledge that as the number of processes increases, so does the size of the information vector. Consequently, this expansion in vector size results in an increased communication overhead required to communicate it among processes. To address this, we introduce a limitation on the range of neighbor information usable by a process during stealing. This means that each process can only access information from a subset of processes, limiting the scope of smart stealing to a local region.
    The bidirectional ring network mesh is configured such that a process can steal from a certain number of processes to its right and left directions, forming a subsystem, as marked in gray in Figure~\ref{fig:info-exp}. The radius ($R$) determines the size of this subsystem. Consequently, the number of elements of the information vector of the subsystem is defined by 
    \begin{equation}
        P_{\text{sub}} = 2R + 1 \text{.}
    \end{equation}
    \noindent Limiting information can potentially increase the incidence of theft, as it serves to balance the entire system. Consequently, intermediary processes must steal tasks from the initial victim to ultimately enable a final process to steal from the intermediaries. This can occur when the initial victim is positioned beyond the radius.
    
    Figure~\ref{fig:info-exp} illustrates the communication among three processes ($p_2$, $p_3$, $p_4$) in a system with eight. Each one has its own information vector with eight positions. However, since we are using a radius of $R = 2$, only $P_{\text{sub}} = 5$.

    If we emphasize the process $p_3$, we can see that the information vector goes from $1$ to $5$. The left part ($[1,2]$) will be sent exclusively by the previous process ($p_2$), and the right part ($[4,5]$) will be sent exclusively by the posterior process ($p_4$). Then, there is an order of exclusivity in writing to each position of the information vector. In other words, for each process $p_i$, its information vector is updated as follows. Process $p_{i-1}$ writes in positions $i-R \le j \le i-1$, the own process $p_i$ writes in position $j= i$, and process $p_{i+1}$ writes in position $i+1 \le j \le i+R$. Then, even though communication occurs asynchronously, no overlapping writes exist. This ensures we can use \textit{Put} operations freely without needing locks that would generate synchronization.

    The information-communication scheme involves sharing two pieces of data from each process: the \textbf{task average runtime} ($t_j$) and the \textbf{total number of tasks} ($n_j$). These pieces of information are stored in the corresponding positions $j$ of the information vector related to each process $i$ being $i-R \le j \le i+R$ and $0 \le i < P$. To ensure the freshness of the information, each position in the vector also has a \textbf{flag} that indicates whether the process needs to communicate the information in position $j$. The flag follows the logic described in Table~\ref{tab:info_flag}, determining when the flag is true. Identifying each position with its own flag is critical so that each process can check if that outdated position is inside the interest region of their neighbors. Otherwise, there is no need to send it.  

    \begin{table} 
    \centering
    \caption{The table describing \textit{when} the flag that describes the information is outdated is set to true and in which position.}
    \label{tab:info_flag}
    \begin{tabular}{c|c|c}
    \textbf{When} & \textbf{Position} & \textbf{Algorithm~\ref{alg:a2ws}}\\ \hline
    $p_i$ updates $t_j$ & Itself ($j=i$) & Line~\ref{l:a2ws:up1} \\
    $p_i$ (thief) steals & Itself, victim & Line~\ref{l:a2ws:st} \\
    $p_i$ (thief) does not steal & victim & Line~\ref{l:a2ws:st} \\
    \begin{tabular}[c]{@{}c@{}}$p_i$ (victim) realized it\\was stolen\end{tabular} & Itself & Line~\ref{l:a2ws:up2} and \ref{l:a2ws:ic}
    \\
    \end{tabular}
    \end{table}
    
    During program execution, the information communication can occur in two moments: $(1)$ right after the program finishes the task, when the program is about to request a new task, and $(2)$ during the task execution if the application allows it.

\subsection{Smart Stealing} \label{subsec:smartStealing}

    Smart stealing aims to adjust the stealing activity using the information collected to avoid excessive steals and improve load-balancing performance. For this, we propose an analytical method to estimate the number of tasks each process requires to ensure all processes complete their tasks simultaneously or as close as possible to that. In a heterogeneous environment, the faster processes should receive more tasks, while the slower processes should receive fewer tasks. 

    Before determining the number of tasks, we first need to know the period necessary to process all tasks should there be no load imbalance, which we call ideal runtime, $t_{ideal}$. Considering the set of all processes as a single system and the totality of tasks, we can devise this using the following equation:
    \begin{equation}
        t_{\text{ideal}} = \frac{\text{Amount of tasks}}{\text{System speed}} = \frac{\sum_{j=0}^{p-1}n_j}{\sum_{j=0}^{p-1}\frac{1}{t_j}} 
        = \frac{N}{T} \text{,}
    \end{equation}
    where $t_j$ denotes the average runtime for the executed tasks and $n_j$ is the total number of tasks of process $j$, including the already executed and available for execution. Then, we determine the ideal number of tasks that a process $i$ needs to execute by multiplying the ideal runtime and the actual speed of the process, given by the inverse of $t_i$, as follows:
    \begin{equation}
        n_{\text{ideal for process $i$}} = \text{Process speed} \times t_{\text{ideal}} = \frac{N}{t_i T} \text{.} 
    \end{equation}    
    Once we have the ideal number of tasks for the process, we subtract the ideal number from the actual number of tasks existing in process $i$ to determine the transfer rate or stealing rate. Thus, we arrive at the following stealing equation:
    \begin{equation} \label{eq:nSteal}
        S_i = \frac{N}{t_i T} - n_i \text{,}    
    \end{equation}
    where $S_i$ is the ideal steal rate for process $i$. When $S_i > 0$, it indicates that process $i$ needs to perform a steal, i.e., acquire tasks from other processes. Conversely, when $S_i < 0$, other processes need to still from process $i$. 

    Equation~\ref{eq:nSteal} considers the complete system information. However, if the neighboring information is limited by the radius $R$, as discussed in Section~\ref{subsec:infoComm}, we need to modify Equation~\ref{eq:nSteal} as follows:
    \begin{equation} \label{eq:nSteal_limited}
        S_i = \frac{\sum_{j=i-R}^{i+R}n_j}{t_i\sum_{j=i-R}^{i+R}\frac{1}{t_j}} - n_i \text{.}    
    \end{equation}

    Equations~\ref{eq:nSteal} and \ref{eq:nSteal_limited} are effective when the steal rate $S_i$ is a rational number, allowing for the stealing of fractions of a task. However, in many cases, task subdivision is not feasible due to the nature of the task or the program performing it. A naive approach to this issue would be rounding the steal rate to the nearest integer. We propose to leverage the gathered information to round $S_i$ in a manner that improves the overall runtime of the program by preventing faster processes from remaining idle during the final stages of execution. In this context, we must determine a rounded steal rate value $D_i$, either $D_i = \lfloor S_i \rfloor$ or $D_i = \lceil S_i \rceil$. 
    
    The choice of $S_i$ that represents the number of tasks the thief $i$ will steal from the victim $j$ results in a runtime $U(S)$ that is the same for the thief and the victim process given by
    \begin{equation}
        U(S) = \frac{n_k + S}{t_k} \text{,}
    \end{equation}
    with $k$ denoting the process index (victim or thief). To choose $D_i$,  we compare the influence of $S_i$ to the thief and victim processes runtime and choose the steal rate that results in the smallest runtime, such that 
    \begin{equation} \label{eq:rounded_rate}
        D_i =
        \begin{cases} 
            \lfloor S_i \rfloor & \text{if } \gamma(\lfloor S_i \rfloor) < \gamma(\lceil S_i \rceil) \text{,}\\
            \lceil S_i \rceil & \text{otherwise,}
        \end{cases}
    \end{equation}
    with 
    \begin{equation}
        \gamma(S) = \text{max} \left (
        U(-S,j),U(S,i)\right ) \text{,}
    \end{equation}
    being $\gamma(S)$, the function that returns the expected runtime for the pair victim and thief for any value of $S$. 

    The stealing rate has an important contribution to the other two aspects of smart stealing. In the absence of information, the state of the task deque becomes the only reliable parameter for determining when stealing will occur. In the context of A2WS, knowing the state of other processes grants it the flexibility to determine when stealing will occur.    With this characteristic, we can develop the preemptive stealing explained in Subsection \ref{subsec:pre_steal}. Another relevant contribution is in the victim selection process that depends on the victim and thief steal rate, which is explained in Section \ref{subsec:vic_sel}.

\subsubsection{Preemptive stealing} \label{subsec:pre_steal}

    Commonly, work-stealing algorithms initiate the stealing process when the deque of the fastest process becomes empty. In our proposal, stealing begins immediately after a process completes its first task. If the thief waits for the other processes to finish their first task, it may introduce a problem for heterogeneous load balancing. In some cases, the performance disparity between the fastest and slowest processes can be significant enough to lead to a situation where the fastest process empties its deque while the slowest process still executes its first task. As a result, the fastest process cannot perform any stealing operations as it waits for the slowest process to complete its first task.

    Therefore, we estimate information about the other processes to enable thieves to steal without waiting for other processes. The number of tasks is relatively straightforward to deduce, as we can assume it remains the same, and all processes know this value because the A2WS distributes the tasks statically just before execution starts. For information regarding the processing speed, we use the elapsed wall time to calculate it. In these cases, we replace $t_i$ in Equation~\ref{eq:nSteal} with the elapsed wall time. 

    With this methodology, when the fastest process completes its first task and no other process has finished any task, the calculation based on the elapsed wall time leads to the assumption that all processes have similar performance. As a result, all processes are expected to require the same number of tasks, and stealing does not occur. Once the fastest process finishes its second task, the processes that have not yet completed their first task will perform at approximately half the speed of the fastest process. Consequently, these processes will have half their tasks available for stealing by the fastest process. This behavior continues as subsequent tasks are completed. However, it is important to note that the final number of tasks available for stealing is limited by the value calculated using Equation~\ref{eq:nSteal}. 

\subsubsection{Victim Selection} \label{subsec:vic_sel}

    To initiate task stealing, the first step is calculating the number of tasks the thief process needs using Equation~\ref{eq:nSteal}. Once we have determined the number of tasks to steal, we must select an appropriate victim. There are two criteria for selecting the victim: \textit{the closest rate} and \textit{in-pair comparison}. Both criteria involve assigning selection probabilities to potential victims. These probabilities help the A2WS algorithm distribute the stealing operations, preventing the overloading of victim processes due to multiple thieves attempting to steal from the same victim.

    The goal of the first criterion is to minimize the overall amount of stealing required. In the best-case scenario, where the victim thief processes have stealing rates with opposite signs but the same absolute rate, only one stealing operation is necessary to load balance the two processes within the subsystem. However, in the worst-case scenario, the thief process steals enough tasks to balance its workload because its rate is lower than that of the victim process, or vice versa.

    Sometimes, the information about the nodes may not be precise due to communication delays or inaccurate information deduction, resulting in an inability to identify potential victims, i.e., $S_j < 0$. To address such situations, the second criterion is utilized. The criterion of in-pair comparison is employed when the subsystem appears balanced, but tasks are still available for stealing. The second criterion also utilizes probability selection but operates on a pair-by-pair basis, calculated as
    \begin{equation} \label{eq:2criterion}
        S_j = \frac{(n_i + n_j)}{t_i/(1/t_i + 1/t_j)} - n_i \text{,}
    \end{equation}
    which we can simplify by
    \begin{equation} \label{eq:2criterion2}
        S_j = \frac{(n_i+n_j)t_j}{t_j+t_i} - n_i.
    \end{equation}
    Equation~\ref{eq:2criterion} adapted the Equation~\ref{eq:nSteal} replacing $N$ and $T$ by the respective value in each pair of victim and thief $n_i + n_j$ and $1/t_i + 1/t_j$. In this case, the thief process $i$ calculates a value of $D_j$ for each potential victim process $j$ within the subsystem, excluding itself, i.e., $j \neq i$. Calculating probabilities on a pair-by-pair basis avoids information inconsistency, ensuring that each pair's weight calculation is based on isolated information. 

\subsection{Asynchronous theft}

    Load balancing methods based on synchronization generate an overhead proportional to the number of processes involved in the execution. Using an extensive system with hundreds of nodes can become impractical due to the excessive synchronization overhead caused by the waiting processes. An asynchronous approach is generally more complex but may generate less overhead. 

    To remove synchronization in A2WS, we propose an asynchronous theft using the Head and Tail System to control access to the deque and Atomic MPI Operations that speed up the communication among processes.    

\subsubsection{Head and Tail System} \label{subsec:h_t}
 
    A2WS uses a head and tail system to control the access to the task deque. Differently from~\cite{Li2013} that allows asynchronous theft by creating a private and shared region on the deque to limit the stealing, we employ MPI One Side Communication (OSC). In MPI OSC, only one process is active in the communication allowing for shared and exclusive locks, which might be used to protect access to memory regions. When an exclusive lock is used, the communication operation is exclusive to the process, and other processes must wait until it finishes. On the other hand, a shared lock allows other communication operations of the same type to occur concurrently. 
    
\begin{figure*}[h]
        \centering
        \begin{subfigure}{\textwidth}
            \centering
            \includegraphics[width=.62\textwidth]{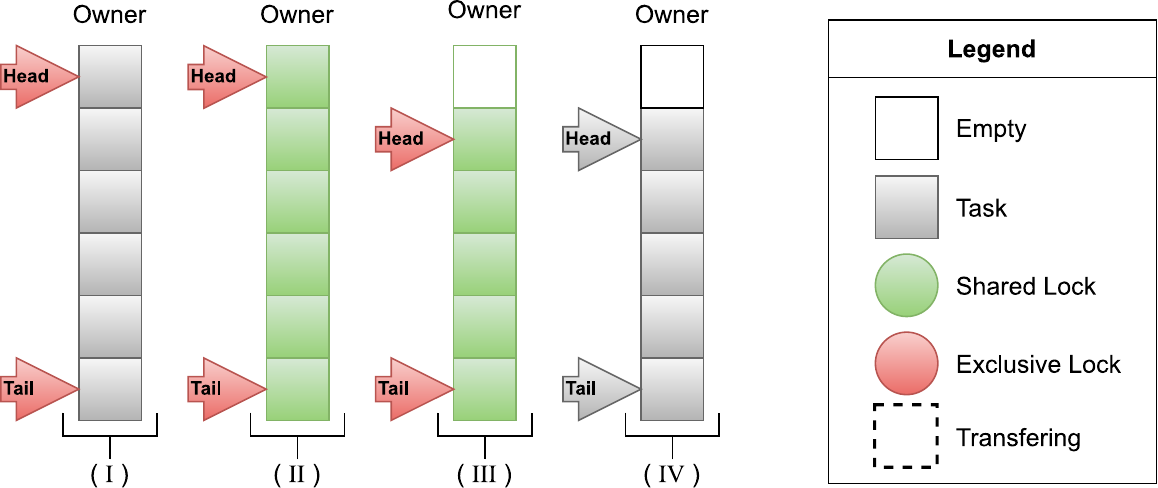}
            \caption{Get Task}
            \label{subfig:get_task}
        \end{subfigure}
        \hfill
        \begin{subfigure}{\textwidth}
            \centering
            \includegraphics[width=\textwidth]{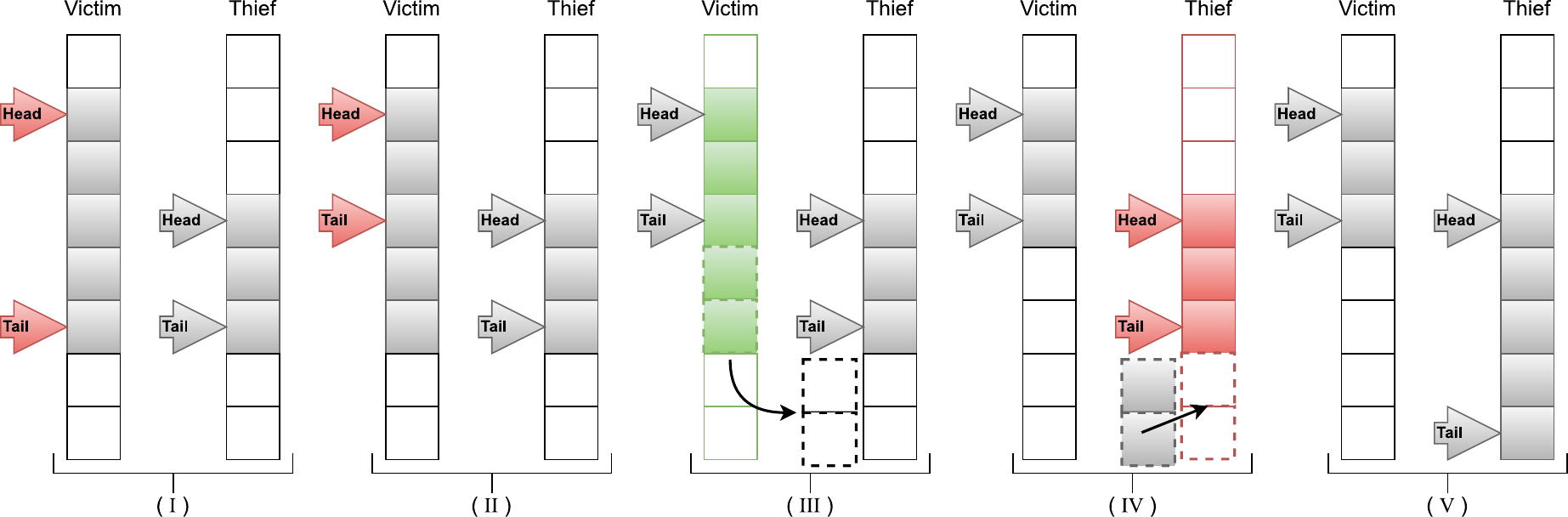}
            \caption{Steal Task}
            \label{subfig:steal_task}
        \end{subfigure}
        \caption{Diagrams of the \textit{get task} and \textit{steal task} A2WS operations demonstrating the steps order of MPI lock and unlock of the \textit{head}, \textit{tail}, and deque. In \ref{subfig:get_task}, we see the \textit{get task} operation perspective of the owner of the task deque in steps I to IV. Whereas, in \ref{subfig:steal_task}, we see the thief operation in steps I to V.}
        \label{fig:work stealing process}
    \end{figure*}

    In the context of A2WS, two crucial asynchronous functions utilize MPI Locks: \textit{Get task} and \textit{Steal task}. The \textit{Get task} function allows the owner of the task deque to obtain a new task for execution. Conversely, the \textit{Steal task}  function is designed to transfer tasks from a victim when necessary. Even though these two functions operate independently, there may be situations where they attempt to access the same process deque simultaneously. Without the assurance of proper locks, this could result in the execution of duplicate tasks or task loss. As shown in Figure~\ref{subfig:get_task}, in step (I), the deque owner locks the head and tail as exclusive, and in step (II), it locks the deque as shared. Then, in step (III), it moves the head. Finally, in step (IV), it unlocks the head, tail, and deque. 

    The \textit{Steal task} involves two processes, the thief (responsible for running the function) and the victim, as shown in Figure~\ref{subfig:steal_task}. In step (I), the thief locks the victim's head and tail as exclusive. In step (II),  it moves the tail and unlocks both in step (III). After this, a temporary vector is created in the thief to transfer the tasks, and the transfer starts. In step (IV), after the transfer is finished, the thief unlocks the victim deque and locks its own deque as exclusive to transfer the tasks from the temporary vector. In the end, in step (V), the thief unlocks its own deque.
    
    Steps (I) and (II) of the \textit{steal task} are designed to prevent the victim from changing the \textit{head} while the thief alters the \textit{tail}. This is crucial to avoid a race condition where the head surpasses the tail position. In such a scenario, the victim would end up executing a task that has been stolen by the thief, resulting in the duplication of task execution. Steps (I) to (II) of Figure~\ref{subfig:get_task} illustrate why the victim is unable to move the head as it attempts to apply an exclusive lock in step (I). On the other hand, Step (III) of Figure~\ref{subfig:steal_task} allows the victim to get new tasks, moving the head, but prevents it from adding new tasks to its deque. In other words, if the victim initiates a steal process while the thief is still in step (II) when the victim reaches step (IV) of the \textit{steal task}, the exclusive lock on the deque blocks it from adding new tasks.

\subsubsection{Atomic Operations}

    While the head and tail approach aims to organize access to the deque and avoid race conditions during multiple access operations on the same deque, recent studies in work-stealing have exploited new MPI functionalities to reduce communication overhead in these operations. \citeay{Cartier2021} proposed a structured atomic operation to minimize the number of stealing operations. They introduced a structure that keeps track of the number of steal attempts, utilizes a shared region to limit the amount of stealing, and restricts the number of steal attempts to half the available tasks. We have chosen to follow a similar approach but provide more flexibility to the thief, maintaining the head and tail.

    We have combined the head and tail into a single structure representing them. When stealing happens, it shifts the tail and returns the head and tail position in a single communication using \textit{MPI\_Get\_accumulate()}. If the thief tries to steal more tasks than the victim has available, it will adjust the amount with the returned positions. In the same case, the victim will detect the stolen tasks when checking the own head and tail, verifying $\text{tail} < \text{head}$ and, after this, classify its deque as empty.
    
    \begin{figure*}
        \centering
        \begin{subfigure}{.4\textwidth}
            \centering
            \includegraphics[width=.8\textwidth]{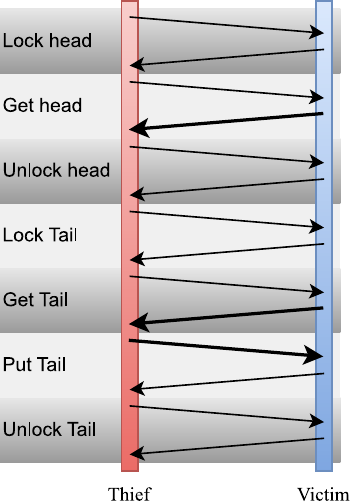}
            \caption{}
            \label{subfig:ht old}
        \end{subfigure}
        \hfill
        \centering
        \includegraphics[width=.18\textwidth]{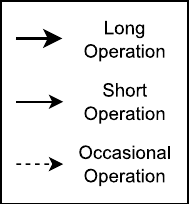}
        \begin{subfigure}{.4\textwidth}
            \centering
            \includegraphics[width=.8\textwidth]{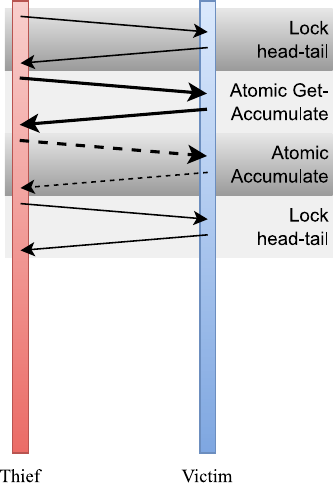}
            \caption{}
            \label{subfig:ht new}
        \end{subfigure}
        \caption{Sequences of operations are performed in the stealing step to obtain information from the victim deque and update the victim tail. It is divided into the raw version (Figure~\ref{subfig:ht old}) with head and tail as separate data, commonly used in work-stealing with MPI, and the adapted version (Figure~\ref{subfig:ht new}) with head and tail as a proposed single data structure.
        The bold arrows represent \textit{long operations} involving data transfer, while the \textit{short operations} do not. The dashed arrows are \textit{Occasional operations} that occur only when there is a disparity between the local and victim information and necessitate stealing adjustments.}
        \label{fig:head tail}
    \end{figure*}

    Figure~\ref{fig:head tail} illustrates the communication operations involved in the stealing process described in Figure~\ref{fig:work stealing process}. Figure~\ref{subfig:ht old} depicts how the head and tail are handled separately. In contrast, Figure~\ref{subfig:ht new} presents the proposed version, where we reduce the number of operations from seven to four. Unlike the approach described in \citep{Cartier2021}, we have an additional MPI operation called the \textit{Atomic Accumulate}. This operation occurs only if the thief's local information about the victim's number of tasks differs from the checked during the \textit{Atomic Get-Accumulate} operation. When such a discrepancy arises, indicating that the thief needs to steal fewer tasks than initially calculated, we must adjust the stealing process and the tail position on the victim. This adjustment is accomplished through the \textit{Atomic Accumulate} operation.
    
\section{Use case: seismic modeling}
\label{sec:wave}

A seismic acquisition is performed in the area of interest. The seismic acquisition process consists of the execution of numerous seismic shots, which are individual experiments conducted across areas spanning several kilometers. On the other hand, modeling involves the numerical simulation of these acquisitions and plays a crucial role in important seismic processing algorithms such as full-waveform inversion (FWI)~\citep{tarantola1984} and reverse-time migration (RTM)~\citep{Baysal1983, Kosloff1983}. These algorithms are computationally intensive due to their reliance on solving complex differential equations that describe the propagation of seismic waves for each individual shot. Furthermore, when modeling acquisitions conducted over vast areas, the overall number of shots can be significantly large.

Seismic modeling consists of numerically simulating the propagation of waves in the Earth's subsurface that happen in the seismic acquisition. Its primary purpose is to estimate the data recorded by sensors placed at known positions, which result from the propagation of seismic waves reflecting within the subsurface in a seismic field. The wave propagation algorithm is a fundamental component of various seismic processing methods, including seismic inversion. To solve the wave equation, the geological model is typically discretized into a finite number of points. Source waves, or seismic shots, are set up at different points close to the surface, simulating real wave propagation. Then, calculations are performed in the discretized model for different and independent shots.

The 3-D acoustic wave equation can be expressed as follows~\citep{Etgen2007}: 

\begin{equation} \label{eq:3dwave}
\frac{1}{c^2 \rho} \frac{\partial^2 u}{\partial t^2} = \nabla \left( \frac{1}{\rho} \nabla u \right).
\end{equation}

\noindent Equation~\ref{eq:3dwave} relates the pressure field $u(x,\,y,\,z,\,t)$ with the density $\rho$ and the velocity $c(x,\,y,\,z)$ at coordinates $x$, $y$, and $z$, and time $t$. 
The steps involved in a computational algorithm that evaluates the wave propagation can be divided into (I) reading the position of the shot; (II) for each position on the model, calculating the wave propagation using Equation~\ref{eq:3dwave}; (III) inserting the source data for the next time step; and (IV) recording the waves that return to the receivers positions in the surface.  

In this work, we employ a simplified version of~\ref{eq:3dwave}, which does not take into account the densities, expressed as:    
\begin{equation} \label{eq:onda}
\frac{\partial^2 u}{\partial x^2} + \frac{\partial^2 u}{\partial y^2} + \frac{\partial^2 u}{\partial z^2} = \frac{1}{c^2}\frac{\partial^2 u}{\partial t^2} + s(t),
\end{equation} where $u(x, \, y, \, z, \, t)$ denotes the acoustic pressure, $c(x, \, y, \, z)$ represents the propagation velocity, and $s(t)$ is the source function at time $t$. For numerically solving Equation~\ref{eq:onda}, we have employed a finite difference method (FDM)~\citep{Moczo2007}.

\section{Results}
\label{sec:results}

    The experiments were conducted on the supercomputer SDumont\footnote{https://sdumont.lncc.br/}, utilizing nodes equipped with 2 x CPU Intel Xeon E5-2695v2 Ivy Bridge, 2.4 GHz, and 64 GB DDR3 RAM. To simulate a heterogeneous machine for our experiments, we took advantage of the heterogeneous-job feature of SLURM~\citep{slurm2003}, a job scheduling and management system that allowed us to allocate different resources to each node. Through this approach, we could limit the number of CPUs used in the tests on specific nodes. For each configuration presented in Table~\ref{tab:tests_config}, we selected groups of nodes with 1, 2, 4, 8, 16, and 24 Cores. We compared A2WS to Learder and Workers (Centralized Dynamic method, LW) and Cyclic Token-based Work-Stealing (CTWS) \cite{Assis2019}. We developed the LW algorithm based on a classical master-slave scheduler where an extra thread of the main process is responsible for distributing all tasks.

    \begin{table}[]
    \centering
    \caption{The test node configurations (C1-C5) are organized in a table. The selected nodes were divided into two parts: one part utilized only 1 and 24 Cores, while the other part used 2, 4, 8, and 16 Cores. This approach was employed to simulate a heterogeneous machine and was facilitated with the help of SLURM's capabilities for managing heterogeneous jobs. For example, Configuration 01 (C1) includes $2$ nodes using $1$ Core, $2$ nodes using $24$ Cores, $1$ node using $2$ Cores, $1$ node using $4$ Cores, $1$ node using $8$ Cores, and $1$ node using $16$ Cores, totalizing $8$ nodes.}
    \begin{tabular}{c|ccccc}
    \textbf{} & \multicolumn{5}{c}{\textbf{Configurations}} \\ \cline{2-6} 
    \textbf{} & \textbf{C1} & \textbf{C2} & \textbf{C3} & \textbf{C4} & \textbf{C5} \\ \hline
    \textbf{1 core node} & 2 & 4 & 8 & 16 & 32 \\
    \textbf{2 core node} & 1 & 2 & 4 & 8 & 16 \\
    \textbf{4 core node} & 1 & 2 & 4 & 8 & 16 \\
    \textbf{8 core node} & 1 & 2 & 4 & 8 & 16 \\
    \textbf{16 core node} & 1 & 2 & 4 & 8 & 16 \\
    \textbf{24 core node} & 2 & 4 & 8 & 16 & 32 \\
    \textbf{Total of nodes} & 8 & 16 & 32 & 64 & 128
    \end{tabular}
    \label{tab:tests_config}
    \end{table}

\begin{figure}
    \centering
    \includegraphics[width=\textwidth]{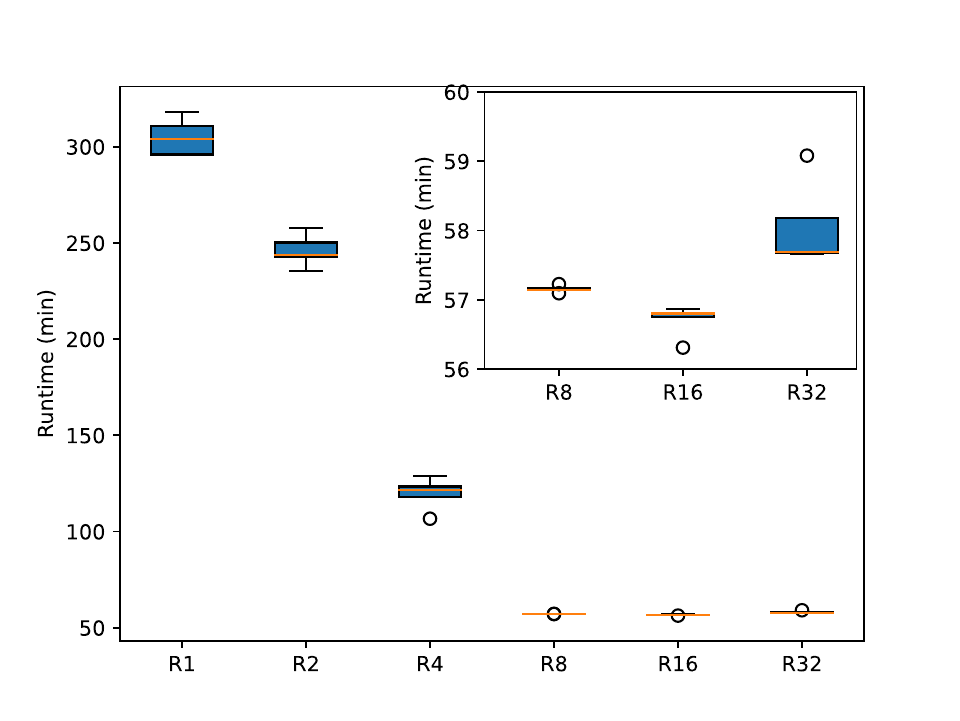}
    \caption{$5$ samples of runtime test in minutes varying the A2WS radius from $R1$ ($R = 1$) to $R32$ ($R = 32$) using the Configuration $04$ of Table~\ref{tab:info_flag} and $3840$ tasks. The subplot represents a subset of the data from $R8$ to $R32$.}
    \label{fig:runtime_radius}
    \end{figure}

    Firstly, we conducted the initial test to fine-tune the Radius parameter ($R$). In this experiment, we utilized \textit{Configuration} $04$ from Table~\ref{tab:tests_config} with a total of $3840$ tasks and $R = \{1, 2, 4, 8, 16, 32\}$. The maximum value of $32$ practically does not have a constraint when using 64 nodes, which is equivalent to global communication. The results of this test are illustrated in Figure~\ref{fig:runtime_radius}. 

    Observe in Figure~\ref{fig:runtime_radius} that the runtime decreases as the radius increases from $R = 1$ to $R = 16$. This behavior can be attributed to the fact that a smaller radius limits the amount of information available to the nodes for achieving a balanced subsystem. Consequently, the impact on the overall system turns the balance insufficient. The subplot in Figure~\ref{fig:runtime_radius} shows the minimum runtime at $R = 16$, which is better than global information exchange with $R = 32$. It is reasonable to assume that for even larger radius values, the communication overhead generated by the size of the information vector could significantly affect negatively the program's execution time. To address this potential issue, we decided to set a fixed radius at $20\%$ of the total number of nodes in the following experiments. 

    The following tests compare the A2WS with other methods in the literature. The reported percentage gain of runtime of the A2WS related to one of the other methods is calculated using the following equation. 
    \begin{equation}
        \text{Gain} = \left( 1 - \frac{\text{A2WS' median runtime}}{\text{Other method's median runtime}} \right) \times 100\% \text{.}
    \end{equation}

    The results presented in Tables~\ref{tab:gain_lw} and \ref{tab:gain_ctws} encompass all configurations listed in Table~\ref{tab:tests_config}, along with task quantities of $480$, $960$, $1920$, $3840$, and $7680$. The displayed values are the median obtained from five runs. Table~\ref{tab:gain_lw} compares the A2WS performance and the LW approach, while Table~\ref{tab:gain_ctws} showcases a comparison with the CTWS approach.
    In both cases, there are configurations where the A2WS experienced significant losses. This usually occurs when a node handles only a few tasks, leading to insufficient time for the A2WS to balance the execution properly. As the processes do not globally verify the status of all other processes before completion, the fastest processes can finish earlier than the optimal time, causing performance degradation in A2WS.

    \begin{table}[]
    \caption{Gain of A2WS in relation to LW using all configurations shown in Table~\ref{tab:tests_config} and the following task amounts: $480$, $960$, $1920$, $3840$, and $7680$. The reported results are the medians of 5 samples.}
    \centering
    \begin{tabular}{c|ccccc}
     & \multicolumn{5}{c}{\textbf{Nodes}} \\
    \multirow{-2}{*}{\textbf{Tasks}} & \textbf{8} & \textbf{16} & \textbf{32} & \textbf{64} & \textbf{128} \\ \hline
    \textbf{480} & \cellcolor[HTML]{86CFAB}16.0\% & \cellcolor[HTML]{B0DFC8}10.5\% & \cellcolor[HTML]{E2F3EB}3.9\% & \cellcolor[HTML]{E67D74}-113.9\% & \cellcolor[HTML]{EA948C}-94.1\% \\
    \textbf{960} & \cellcolor[HTML]{7ECBA5}17.1\% & \cellcolor[HTML]{BFE5D3}8.5\% & \cellcolor[HTML]{B4E1CB}10.0\% & \cellcolor[HTML]{57BB8A}22.2\% & \cellcolor[HTML]{E67C73}-115.3\% \\
    \textbf{1920} & \cellcolor[HTML]{8FD2B1}14.8\% & \cellcolor[HTML]{BFE5D3}8.5\% & \cellcolor[HTML]{EEF8F3}2.3\% & \cellcolor[HTML]{B0DFC8}10.5\% & \cellcolor[HTML]{FEFEFE}-0.0\% \\
    \textbf{3840} & \cellcolor[HTML]{A2DABE}12.3\% & \cellcolor[HTML]{C3E7D6}7.9\% & \cellcolor[HTML]{C7E9D8}7.4\% & \cellcolor[HTML]{D3EEE1}5.8\% & \cellcolor[HTML]{B3E0CA}10.1\% \\
    \textbf{7680} & \cellcolor[HTML]{8FD2B1}14.9\% & \cellcolor[HTML]{C6E8D8}7.5\% & \cellcolor[HTML]{D3EEE1}5.9\% & \cellcolor[HTML]{D1ECDF}6.2\% & \cellcolor[HTML]{D2EDE0}6.0\%
    \end{tabular}
    \label{tab:gain_lw}
    \end{table}

    \begin{table}[]
    \caption{Gain of A2WS in relation to CTWS using all configurations shown in Table~\ref{tab:tests_config} and the following task amounts: $480$, $960$, $1920$, $3840$, and $7680$. The reported results are the medians of 5 samples.}
    \centering
    \begin{tabular}{c|ccccc}
      & \multicolumn{5}{c}{\textbf{Nodes}} \\
    \multirow{-2}{*}{\textbf{Tasks}} & \textbf{8} & \textbf{16} & \textbf{32} & \textbf{64} & \textbf{128} \\ \hline
    \textbf{480} & \cellcolor[HTML]{D2EDE0}5.88\% & \cellcolor[HTML]{CCEBDC}6.73\% & \cellcolor[HTML]{E3F4EB}3.75\% & \cellcolor[HTML]{E67C73}-116.27\% & \cellcolor[HTML]{EA938C}-95.01\% \\
    \textbf{960} & \cellcolor[HTML]{F1FAF6}1.84\% & \cellcolor[HTML]{C7E9D8}7.31\% & \cellcolor[HTML]{B7E2CD}9.43\% & \cellcolor[HTML]{57BB8A}21.88\% & \cellcolor[HTML]{E67C73}-116.04\% \\
    \textbf{1920} & \cellcolor[HTML]{FAFDFC}0.70\% & \cellcolor[HTML]{EFF9F4}2.14\% & \cellcolor[HTML]{EFF9F4}2.20\% & \cellcolor[HTML]{B2E0CA}10.08\% & \cellcolor[HTML]{FEFEFE}-0.03\% \\
    \textbf{3840} & \cellcolor[HTML]{FEFDFD}-0.96\% & \cellcolor[HTML]{E7F5EE}3.25\% & \cellcolor[HTML]{EBF7F1}2.73\% & \cellcolor[HTML]{D2EDE0}5.96\% & \cellcolor[HTML]{B2E0C9}10.15\% \\
    \textbf{7680} & \cellcolor[HTML]{F0F9F5}2.06\% & \cellcolor[HTML]{F5FBF8}1.40\% & \cellcolor[HTML]{E9F7F0}2.89\% & \cellcolor[HTML]{D0ECDF}6.14\% & \cellcolor[HTML]{D2EDE0}5.89\%
    \end{tabular}
    \label{tab:gain_ctws}
    \end{table}

    \begin{figure}
    \centering
        \begin{subfigure}[b]{0.5\textwidth}
        \includegraphics[width=\textwidth]{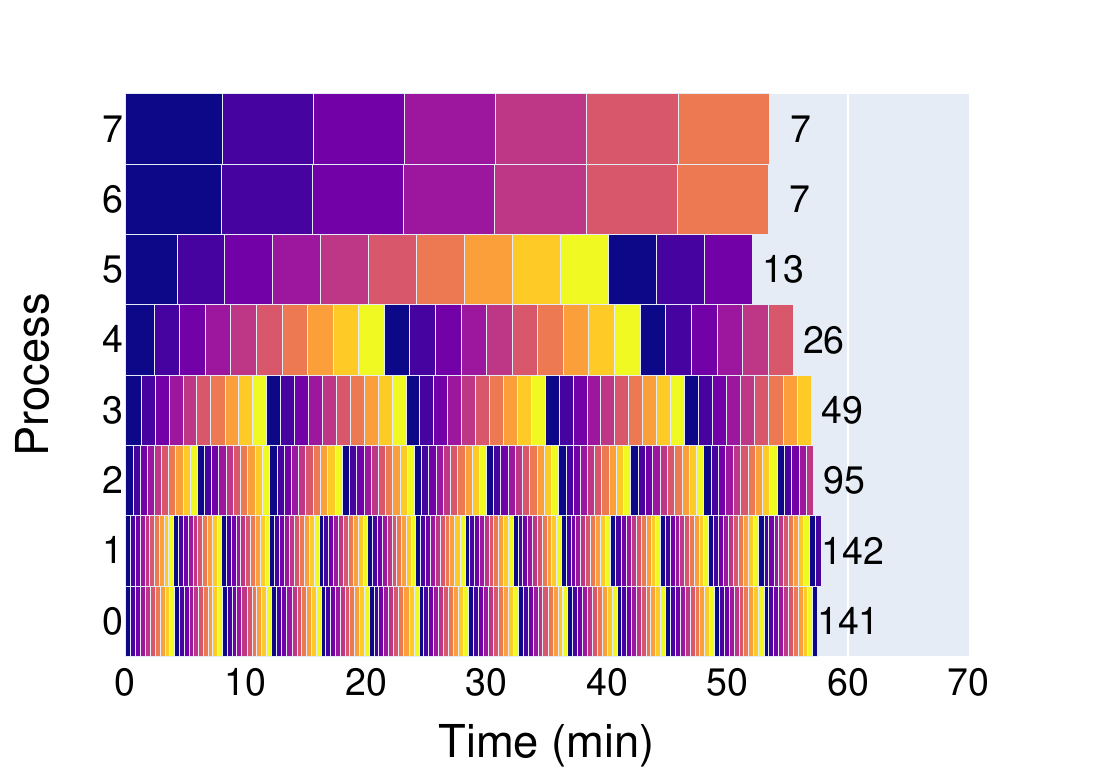}
        \caption{A2WS task runtime}
        \label{subfig:8_tasks_a2ws}
        \end{subfigure}
        \begin{subfigure}[b]{0.5\textwidth}
        \includegraphics[width=\textwidth]{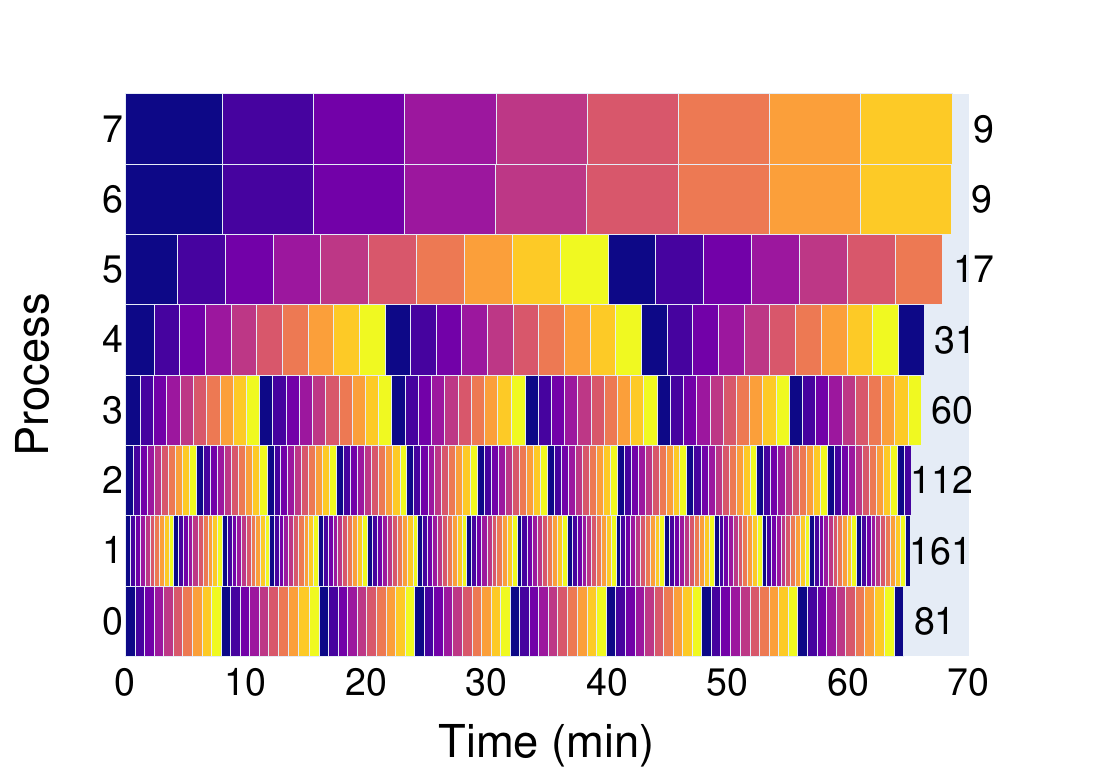}
        \caption{LW task runtime}  
        \label{subfig:8_tasks_lw}
        \end{subfigure}
        \begin{subfigure}[b]{0.5\textwidth}
        \includegraphics[width=\textwidth]{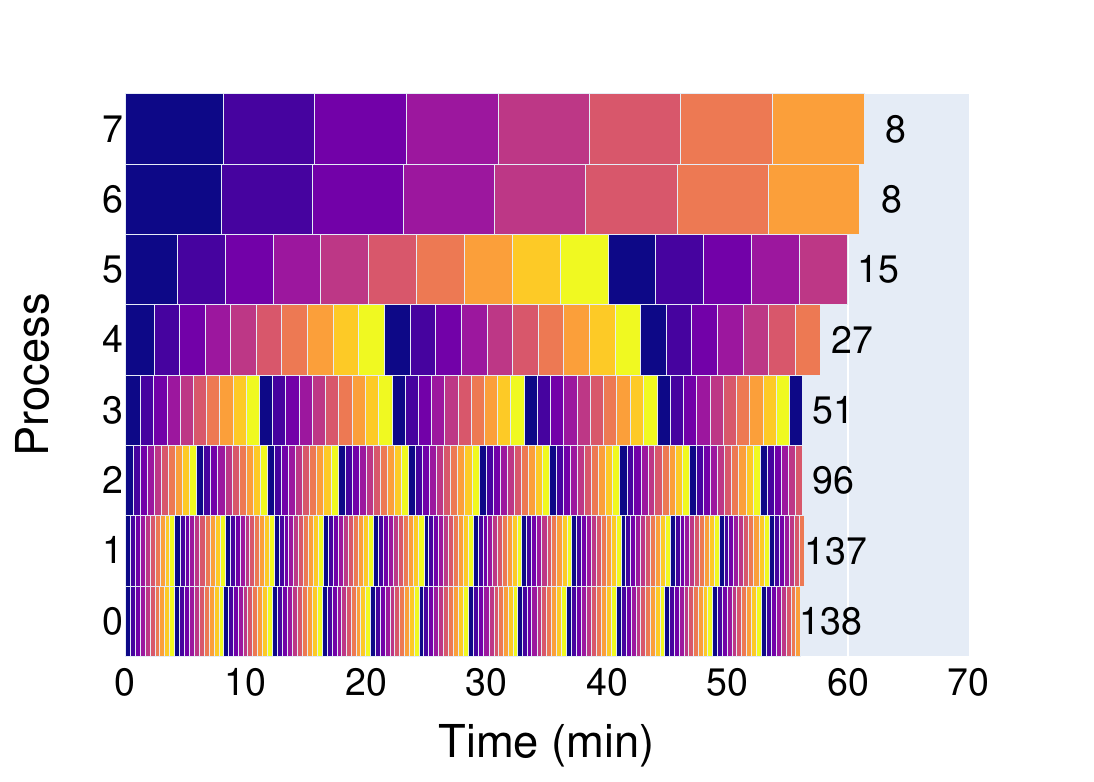}
        \caption{CTWS task runtime}
        \label{subfig:8_tasks_ctws}
        \end{subfigure}
    \caption{Runtime (in minutes) per task of A2WS, CTWS and LW using \textit{Configuration} $1$ in Table~\ref{tab:tests_config} and $480$ tasks. At the end of each line is described the number of tasks executed.}
    \label{fig:8_tasks_runtime}
    \end{figure}

    The results shown in Figure~\ref{fig:8_tasks_runtime} were generated using configuration $C1$ and a total of $3840$ tasks. In this setup, we executed the A2WS, LW, and CTWS and measured the runtime for each task in each process. At the end of each line, we provide the number of tasks executed by each process. If we compare Figure~\ref{subfig:8_tasks_a2ws} and \ref{subfig:8_tasks_ctws}, we can see the task runtime of all processes is similar. The difference is in the slowest processes, $6$ and $7$. These processes execute one task more in the CTWS compared with A2WS, and this task can have a significant impact on the overall runtime, as we can see in the test with $8$ nodes and $480$ tasks where the A2WS gain is $5.88\%$ in Table~\ref{tab:gain_ctws}. This happens because the A2WS prioritizes the fastest processes, while the CTWS cannot do this because it does not account for the processes' speed. This effect is most evident in the CTWS but also happens in the LW, as shown in Figure~\ref{subfig:8_tasks_lw}, where the slowest processes ran $9$ tasks.

    In Table~\ref{tab:gain_lw}, we can observe that A2WS' gains with $8$ nodes are high on average, varying between $12.3\%$ and $17.1\%$. This is due to a disadvantage of LW, which experiences a performance drop on its main node where an extra thread is dedicated to task distribution. We can observe this impact in Figure~\ref{subfig:8_tasks_lw}, where Process $O$ exhibits a higher average task execution time compared to the same process in Figures~\ref{subfig:8_tasks_a2ws} and \ref{subfig:8_tasks_ctws}. With this performance degradation in Process $0$, tasks are redistributed to other processes, increasing the overall program runtime.

    As we increase the number of nodes, the influence of the additional thread becomes less pronounced, resulting in an advantage for LW as opposed to A2WS. This phenomenon becomes evident as the gains for A2WS tend to decrease until reaching a minimum point, after which they increase again. Typically, this minimum point occurs at around $32$ nodes in most scenarios. This observation can be attributed to A2WS functioning as a decentralized method. Consequently, the primary node in LW, which assumes the role of centralizing task distribution, becomes increasingly overloaded as the number of nodes is raised.

    Similarly, the performance gain of A2WS increases compared to CTWS as the number of nodes rises. This shift is notable, for instance, in the transition from $-0.96\%$ to $10.15\%$ when employing $3840$ tasks in Table~\ref{tab:gain_ctws}. This phenomenon is connected to the use of a single token in CTWS, which regulates the stealing and transmitting of global information among nodes. As the number of nodes grows, the size of the token also expands and accommodates more nodes in circulation. These two factors combine to extend the time required for the token to complete its circulation. Consequently, the chance of the theft process experiencing prolonged waiting times before initiating stealing also increases, as it must await the arrival of the token. In contrast, the fully asynchronous nature of A2WS prevents this behavior from occurring.

    The smart stealing mechanism of A2WS plays a crucial role in enhancing task redistribution. This mechanism prioritizes faster processes over slower ones, which has a significant impact on the runtime during the final stages of execution. As depicted in Figure~\ref{fig:8_tasks_runtime}, when a slow process empties its deque while others still have tasks to execute, it cannot steal due to the constraints imposed by the smart stealing rules. Simultaneously, the asynchronicity of A2WS enhances its scalability, as evidenced by the increased gains in A2WS performance when the number of processes is increased, as shown in Tables~\ref{tab:gain_lw} and \ref{tab:gain_ctws}. Summarizing the results, A2WS is more advantageous than the other methods for larger numbers of tasks per node combined with higher numbers of nodes. Jobs with only one of these characteristics are less suitable for our method.

\section{Related Work}
\label{sec:related}

Many papers primarily address load-balancing in environments characterized by heterogeneity where the CPU is coupled with a specific-purpose processor, such as a GPU~\citep{Gregg2011, Vet2013, Kaleem2014, Lima2015, Shulga2016, Martinez2015}, a DSP~\citep{Kumar2015}, or where the architecture are reconfigured, as an FPGA~\citep{Zhang2022}. On the other hand, there is a group of works that applies scheduling programs for clusters or data centers, aiming to assign them to the most suitable machines to reduce runtime~\citep{Im2016A, Im2016B, Delimitrou2013} or minimize energy costs~\citep{Shaolei2014}. Our objective is not to limit the program to a specific architecture but to seamlessly integrate various architectures into its execution. This approach eliminates the need for users to spend time selecting optimal nodes, allowing them to leverage both older and slower nodes without hassle.
    
\citeay{Khaitan2013} uses master-slave scheduling, where the master proactively distributes tasks to the slaves to avoid the slaves waiting for new tasks. The master-slave strategy offers a suitable solution for scheduling tasks on demand. On the other hand,~\cite{Tesser2019} proposed utilizing information from previous time steps of the application, using the AMPI/Charm++ application to schedule tasks. However, the scheduling decision remains centralized. In both cases, the centralized aspect tends to overload the process that handles all the communication when the number of nodes scales. This is unlike A2WS, which was designed to be decentralized and scalable.
    
To address scalability issues caused by centralized decisions, some authors have proposed decentralized scheduling methods. \citeay{Zheng2011} proposed a periodic scheduling approach primarily aimed at scaling in large systems. Meanwhile,~\cite{Sharma2014} tested dynamic scheduling in a heterogeneous multi-node cluster without prior knowledge of the system. Both methods operate in multi-node heterogeneous environments, similar to the ones in our work. However, they utilize synchronous steps for task rescheduling. In general, synchronization introduces overhead since processes need to pause and communicate, which can reduce overall program performance. 
    
Among the dynamic scheduling strategies, there is a well-known approach called \emph{work-stealing} (WS)~\citep{Blumofe1999}. It consists of a decentralized scheduling method where faster processes steal tasks from slower ones. Work-stealing is widely used in the literature. \citeay{Li2013} utilized a head and tail system on the task queue with private and shared regions, where each process consumes its own head and private tasks, while the tail and shared tasks are available for theft. This approach for the queue has proven to be effective in making theft completely asynchronous. In our work, we present a similar concept for the queue, utilizing the MPI lock system. Instead of dividing the queue into regions, we classify the head and tail queue as shared or exclusive, enabling the thief to transfer any required number of tasks. On the other hand, \citeay{Kumar2016} introduced a load-aware WS policy to reduce the number of failed steals, where thieves only return to try to steal a victim that returns its requests. However, it uses no victim performance information to improve victim selection.
    
Regarding scalability, \citeay{Dinan2009} discussed the scalability of WS in a distributed memory system, while \citeay{Vishnu2015} introduced WS using MPI one-sided communication for machine learning and data mining algorithms. Both works employ random victim selection without proposing methodologies to improve the organization of stealing. Furthermore, \citeay{Cartier2021} focused on decreasing WS communication by utilizing Remote Direct Memory Access (RDMA) atomic operations, which inspired some aspects incorporated into our work adapting in the context of sharing information, a context that was not addressed in \citeauthor{Cartier2021} work. Additionally,~\cite{Gast2021} proposes a study of the impact of communication latency on classical Work Stealing and provides a theoretical analysis and simulation to predict performance under different conditions. 

The proposal of \citeay{Wenjie2020} combines adaptive persistence-based load balance with WS for single-node environments. \citeay{Nakashima2019} extend a WS framework with priority-based and weight-based stealing for shared memory environments. \citeay{Reitz2023} investigates a hybrid scheme combining inter-process work-stealing with intra-process work-sharing using a pure WS with random victim selection and offload of half amount of tasks.  The probabilistic guards~\citep{Yoritaka2019} aim to prevent the stealing of tasks small enough that copying data is more costly than the gain from transferring. \citeay{Yang2022} proposes a hybrid load-balancing approach with an additional coordinator thread for verification, which can introduce computational overhead.  \citeay{Freitas2021} employs global information to classify the tasks, not the environment, that are most suitable for heterogeneous tasks in a homogeneous environment. \citeay{Chung2023} uses reactive and proactive work-stealing with a dedicated thread for task characterization used in machine learning, which can generate overhead due to periodic status checking on all processes. Our approach improves upon this by eliminating the need for an extra thread, and the information exchange tries to minimize overall communication.

\begin{table*}[]
\caption{Most recent articles in the literature review on work-stealing. The proposed work-stealing method benefits from MPI one-sided communication to further reduce communication overhead. It uses global information and adapts the stealing to improve the scheduling without needing a new thread to collect information and is applied to the seismic algorithm.}
\resizebox{\textwidth}{!}{
\begin{tabular}{|c|c|c|c|c|c|c|c|}
\hline
 &
  \cellcolor[HTML]{FFFFFF}\textbf{\begin{tabular}[c]{@{}c@{}}MPI\\ OSC\end{tabular}} &
  \cellcolor[HTML]{FFFFFF}\textbf{Async.} &
  \cellcolor[HTML]{FFFFFF}\textbf{\begin{tabular}[c]{@{}c@{}}Global Tasks \\ Information\end{tabular}} &
  \cellcolor[HTML]{FFFFFF}\textbf{Adaptive} &
  \cellcolor[HTML]{FFFFFF}\textbf{Preemptive\footnote{The stealing happen before the queue is empty.}} &
  \cellcolor[HTML]{FFFFFF}\textbf{\begin{tabular}[c]{@{}c@{}}Extra \\ Thread\end{tabular}} &
  \cellcolor[HTML]{FFFFFF}\textbf{\begin{tabular}[c]{@{}c@{}}Heterogeneous \\ Environment\end{tabular}} \\ \hline
\textbf{Our purpose} & Yes & Yes & Yes & Yes & Yes & No  & Yes \\ \hline
\citeay{Thanh2023} & Yes & Yes & Yes & Yes & Yes & Yes & No  \\ \hline
\citeay{Reitz2023} & No  & Yes & No  & No  & No  & Yes & No  \\ \hline
\citeay{Yang2022} & No  & Yes & No  & No  & No  & Yes & No  \\ \hline
\citeay{Freitas2021} & No  & No & Yes  & Yes  & Yes & No & No  \\ \hline
\citeay{Cartier2021} & Yes & Yes & No  & No  & No  & No  & No  \\ \hline
\citeay{Gast2021} & No  & No  & No  & No  & No  & No  & No  \\ \hline
\citeay{Wenjie2020} & No  & No  & Yes & Yes & No  & No  & No  \\ \hline
\citeay{Assis2019} & Yes & Yes & Yes & No  & No  & No  & No \\ \hline
\citeay{Nakashima2019} & No  & Yes & No  & No  & No  & No  & No  \\ \hline
\citeay{Yoritaka2019} & No  & Yes & No  & No  & No  & No  & No  \\ \hline
\end{tabular}
}
\end{table*}

In contrast, the work of \citeay{Assis2019} introduces and demonstrates the benefits of Cyclic Token-based Work-Stealing (CTWS) for executing programs with heterogeneous behavior. The CTWS approach incorporates three aspects of WS: asynchronous communication, one-sided communication with MPI RMA, and global load information. To manage information and control stealing, CTWS utilizes a cyclic token. Only the process with the token can perform stealing, thereby avoiding deadlocks and race conditions. While the token mechanism works well with a small number of processes, as the number of processes increases, the overhead of waiting for the token also increases. On the other hand, CTWS utilizes the global information of task numbers to determine stealing victims. However, this approach does not consider the performance of each process, and the number of tasks stolen is always half of the available tasks. Consequently, it leads to excessive stealing attempts and redundant stealing, where a previous victim becomes a thief and steals tasks already stolen. This excessive stealing generates additional communication overhead.

\section{Conclusions}
\label{sec:conclusions}

We have presented the scheduler strategy named adaptive asynchronous work-stealing (A2WS). A2WS collects information about the computer node's performance and adapts aspects of the work-stealing, such as the stealing amount and the victim selection, improving the load balance. Each process shares workload information using non-blocking one-side communication through MPI atomic operations. Each node only communicates its information to a limited number of nodes to avoid excessive global communication, making the algorithm more scalable. This information is used to calculate an ideal number of tasks for each node, determine the number of tasks that each process needs to steal or give away and select the best victim for each stealing. The stealing starts just after the first task is completed, giving time to each process to balance its tasks without waiting for its deque to empty.

We compared A2WS with two other schedulers conducting tests on a seismic modeling algorithm to analyze their performance. The results show that the A2WS provides better task distribution among nodes, prioritizing the fastest node and leading to a decrease in overall runtime. Additionally, we observed that the performance gain of A2WS over CTWS increases as the number of nodes scales, reaching a significant gain of $10.15\%$ with $128$ nodes. This suggests that using A2WS, with its tokenless and fully asynchronous approach, is advantageous compared to CTWS for larger numbers of nodes.

Furthermore, we evaluated the performance of A2WS against LW and found that the A2WS gains reach a minimum point at $32$ nodes and then begin to scale up again. With $128$ nodes, it achieves a gain of $10.1\%$. This indicates the impact of increasing communication overhead on the main node processes and highlights the advantage of using a decentralized method like A2WS. Overall, our analysis demonstrates that A2WS outperforms CTWS regarding task distribution and provides advantages over LW by efficiently managing communication overhead in large-scale systems.

So far, the A2WS has demonstrated performance improvements compared to other implementations. Nonetheless, we still face limitations regarding the types of tasks employed. Typically, these tasks have been homogeneous and executed in heterogeneous environments. The next phase of our work involves categorizing the tasks based on their time significance and adjusting the stealing equation to consider these task weights.

\section*{CRediT authorship contribution statement}

\textbf{João B. Fernandes:} Conceptualization, Data curation, Formal analysis, Investigation, Methodology, Software, Validation, Visualization, Roles/Writing - original draft;
\textbf{Italo A. S Assis:} Methodology, Supervision, Writing - review \& editing;
\textbf{Idalms M. S. Martins:} Writing - review \& editing;
\textbf{Tiago Barros:} Writing - review \& editing;
\textbf{Samuel Xavier-de-Souza:} Conceptualization, Funding acquisition, Methodology, Project administration, Resources, Supervision, Roles/Writing - original draft, Writing - review \& editing.




\section*{Code availability}

    The complete A2WS code is stored in the GitLab repository (\nolinkurl{https://gitlab.com/lappsufrn/a2ws}).

\section*{Acknowledgments}

The authors gratefully acknowledge support from Shell Brazil through the ``New computationally scalable methodologies for target-oriented 4D seismic in pre-salt reservoirs'' project at the Federal University of Rio Grande do Norte (UFRN) and the strategic importance of the support given by the Brazilian National Agency for Petroleum, Natural Gas, and Biofuels (ANP) through the R\&D levy regulation. 
The authors are also thankful for SDumont (High-Performance Computing Center at the National Laboratory of Scientific Computing, LNCC).

\bibliographystyle{elsarticle-harv} 
\bibliography{elsarticle} 

\end{document}